\documentclass[12pt]{article}
\usepackage [dvips]{graphicx} 

\newcommand{\be}{\begin{equation}}

\newcommand{\ee}{\end{equation}}

\newcommand{\ba}{\begin{array}}

\newcommand{\ea}{\end{array}}

\newcommand{\bc}{\begin{center}}

\newcommand{\ec}{\end{center}}

\newcommand{\disregard}[1]{{}}

\begin{document}

\title{Variational HFB Equations in the Thomas-Fermi Limit for Ultracold Trapped Gases}

\author{ M. Benarous \\
Laboratory for Theoretical Physics and Material Physics \\
Faculty of Sciences and Engineering Sciences \\
Hassiba Benbouali University of Chlef  \\
B.P. 151, 02000 Chlef, Algeria. }

\date{\today}

\maketitle

\begin{abstract}

We derive variationally the HFB equations for a trapped self-interacting Bose gas at finite temperature. In the
Thomas-Fermi limit, we obtain simple expressions for the condensate, the non condensate and the anomalous densities. Their behavior in terms of the condensate fraction meets qualitatively the experimental data. In particular, the non condensate and the anomalous densities are peaked at the center of the trap and not at the edges as predicted by the self-consistent HFB calculations.

\end{abstract}

PACS: 05.30.Jp, 11.15.Tk, 32.80.Pj

\newpage

\setcounter{section}{0} 
\setcounter{equation}{0}

\begin{center}
{}
\end{center}

\section{Introduction}

Since the discovery of the Bose-Einstein condensation, a great effort was devoted by researchers all around the world in order to understand and predict the condensate properties. The main tools, beside the Monte-Carlo calculations\cite{KR}, were the Bogoliubov\cite{BG}, the Popov\cite{PO}, the Beliaev\cite{BL,GR} and the Hartree-Fock-Bogoliubov\cite{HF,HU00,KT97,ST} approximations. These approximations all adopt some simplifying assumptions about the various order parameters such as the condensate density $n_c\equiv |\Phi|^2$, the 
non condensed density $\tilde{n}$ and the anomalous density $\tilde{m}$. 

In a previous paper\cite{BM05}, we rely on the time-dependent variational principle of Balian and V\'en\'eroni\cite{BV}, which allows one to overcome some of the restrictions related to the various approximation schemes. We obtained a set of three coupled dynamical equations, which we called ``Time-Dependent Hartree-Fock-Bogoliubov'' (TDHFB) equations, governing the evolution of the order parameter $\Phi $, $\tilde{n}$ and $\tilde{m}$. They were shown to generalize in a consistent way the Gross-Pitaevskii equation\cite{GP}.

In this paper, we focus on the static solutions of our TDHFB equations both at zero and finite temperature in the Thomas-Fermi (TF) regime. The interest is evident, since there remain many unanswered questions such as
 the general dependence of the density profiles on the temperature and on the interaction strength and the effect of the interactions on the critical temperature. More precisely, recent experiments are raising challenging 
questions about the structure of the thermal cloud and its backeffects on the condensate\cite{gerbier} and the role of the anomalous density. 
Two key questions are the observed compression effect of the thermal cloud on the condensate with increasing temperature and the fact that the thermal cloud and the anomalous average are peaked at the center of the trap. However, the full self-consistent HFB calculations\cite{HU00} not only miss the first effect, but also predict that the non condensate and the anomalous densities are rather peaked at the edge of the condensate. One possible reason is the small atom numbers used in the above calculations. However, going to large atom numbers is clearly a quite formidable task in these methods. 

The goal of the present paper is to show that our equations reproduce correctly the compression effect and the shift of the maximum of the densities toward the borders of the trap when the number of atoms is large enough. Hence, the approximations we are actually using (mean field and Thomas-Fermi), when they are valid, retain the most important qualitative features without destroying the underlying physics. This may provide a simple enough tool which can be considered as a starting point for a more elaborate treatment.

After recalling the main steps that we have used to derive the variational TDHFB equations, we
 present in section 2 the static solutions and discuss their properties at zero temperature. 
At finite temperature, the equations are much more involved and require a careful analysis. 
In the TF limit, we present a simple method which allows for a self-consistent determination 
of the various density profiles as well as some other static properties of the condensate such as 
the chemical potential and the condensate radius. Indeed, the TF approximation obviously provides 
simple enough analytical expressions since it amounts to neglecting the second order derivatives thus 
yielding algebraic equations instead of partial differential equations. This is the main advantage 
of our method which yields quite natural results without having to handle highly non-linear self-consistent equations.

In section 3, we present the results of our calculations. We plot first the condensate radius and the central density as functions of the condensate fraction and note in particular the compression effect of the condensate due 
to the thermal cloud. Moreover, we discuss the TF profile obtained for the condensate density even at low condensate fraction. The non condensate density is also plotted for a wide range of condensate fraction and 
shows a good qualitative agreement with recent experiments. Finally, the anomalous density, although not yet measured experimentally, is shown to behave in a quite intuitive way. 

Some concluding remarks and improvements of our method are given at the end of the paper.

\setcounter{equation}{0}
\begin{center}
{}
\end{center}

\section{The Variational HFB Equations and their Thomas-Fermi Limit}
The general TDHFB equations were derived in ref.\cite{BM05} for a grand canonical Hamiltonian of trapped bosons with mass $m$ and quartic self-interactions (with coupling constant $g$): 

\begin{equation}
H=\int_{{\bf r}}a^{+}({\bf r})\left[ -{\frac{\hbar ^{2}}{2m}}\Delta +V_{{\rm 
{ext}}}({\bf r})-\mu \right] a({\bf r})+{\frac{g}{2}}\int_{{\bf r}}a^{+}(
{\bf r})a^{+}({\bf r})a({\bf r})a({\bf r}).  
\label{eq1}
\end{equation}
The quantity $V_{{\rm {ext}}}({\bf r})$ is the trapping potential and $\mu $ is the chemical potential. These equations read: 

\begin{equation}
\begin{array}{rl}
i\hbar \dot{\Phi} & =\left( -{\frac{\hbar ^{2}}{2m}}\Delta +V_{{\rm {ext}}
}-\mu +gn_{c}+2g\tilde{n}\right) \Phi +g\tilde{m}\Phi ^{\ast }, \\ 
i\hbar \dot{\tilde{n}} & =g\left( \tilde{m}^{\ast }\Phi ^{2}-\tilde{m}{\Phi
^{\ast }}^{2}\right) , \\ 
i\hbar \dot{\tilde{m}} & ={\frac{g}{V}}(2V\tilde{n}+1)\Phi ^{2}+4\left( -{\frac{\hbar
^{2}}{2m}}\Delta +V_{{\rm {ext}}}-\mu +2gn+{\frac{g}{4V}}(2V\tilde{n}
+1)\right) \tilde{m},
\end{array}
\label{eq2}
\end{equation}
where $V$ is the volume of the gas. In Eqs.(\ref{eq2}), $\Phi$ is the order parameter, $n_{c}$ the condensate density ($n_{c}=|\Phi |^{2}$), $\tilde{n}$ the non-condensed density (or thermal cloud) and $\tilde{m}$ is the anomalous 
density. The quantity $n\equiv {n_{c}}+\tilde{n}$ is the total density.

The TDHFB equations with a general Hamiltonian $H$ were derived in \cite{BF99}. 
The properties discussed here and in \cite{BM05} were established there in their general forms. These equations were obtained using the Balian-V\'{e}n\'{e}roni variational principle\cite{BV}, with a gaussian trial density operator (that is, an exponential operator of a quadratic form) in the creation and annihilation operators. The result was a set of coupled evolution equations for the expectation values 
$\langle a\rangle$, $\langle a^{+}a\rangle -\langle a^{+}\rangle\langle a\rangle$ and 
$\langle aa\rangle -|\langle a\rangle |^{2}$. When one identifies these quantities respectively with the order parameter $\Phi$, the non-condensed density $\tilde{n}$ and the anomalous density $\tilde{m}$, and when one restricts $H$ to the class (\ref{eq1}), the equations (\ref{eq2}) follow.
 
The TDHFB equations couple in a consistent and closed way the three densities. They should in principle yield the general time, space and temperature dependence of the various densities. Furthermore, they obviously constitute a natural extension of the Gross-Pitaevskii equation\cite{GP}. They are not only energy and number conserving, but also satisfy the Hugenholtz-Pines theorem (see below) which leads to a gapless excitation spectrum in the uniform limit. The two last equations in (\ref{eq2}) are not totally independent since $\tilde{n}$ and $\tilde{m}$ are related
 by the \textquotedblright unitarity\textquotedblright\ relation\cite{BM05}:

\begin{equation}
I=\left( 1+2V\tilde{n}\right) ^{2}-\left( 2V|\tilde{m}|\right) ^{2},
\label{eq2t}
\end{equation}
where the Heisenberg parameter $I$ (which is always $\geq 1$) is a measure of the temperature, the lower limit ($I=1$) being the zero temperature case. For instance, for a thermal distribution at equilibrium, $I$ writes as 
$I=\coth^{2}{(\hbar \omega _{0}/2{k_{B}}T)}$, where $\omega _{0}$ is the average frequency of the trapping field\cite{BM05} \footnote{In fact, one can show that for a system of energy $E$, $\sqrt{I}=1+2\,f_{B}(E)$, where $f_{B}$ is the Bose-Einstein distribution.}. We therefore see that upon replacing $\tilde{n}$ by its expression given in 
(\ref{eq2t}), the temperature appears explicitly in the equations. 

It is to be mentioned that the TDHFB equations have also been derived by other groups using different variational formulations\cite{CHER03, Pr04}. For instance, the authors of the first reference have used a gaussian trial wave 
functional and have obtained a set of equations very similar to ours. However, their discussion is purely formal and does not address for instance the question of validity of their approximation.

The static solutions, which are the object of our study in this work, are obtained by setting to zero the right hand sides of (\ref{eq2}). At zero temperature, the standard TF limit\cite{SV} amounts to neglecting the kinetic
(or $\Delta $) term in the Gross-Pitaevskii equation. This is particularly satisfied for strong interacting regimes or large atom numbers. At finite temperature and below the transition, since there are two phases (condensed 
and non condensed) which coexist, one has to provide a complementary recipe for what we shall call the finite temperature TF limit. First, neglecting the kinetic energy of the condensate remains a justifiable approximation since the atoms are slowed down in order to obtain condensation. On the other hand, $\tilde{m}$ is believed to be an extremely small and slowly varying function whatever the temperature is (recall that it describes the correlations between the condensed and non-condensed phases). Hence, one may in a first approximation safely neglect 
$\Delta \tilde{m}$. Heuristically, one may argue that, since the equations for $n_c$ and $\tilde{m}$ contain almost comparable operators, $h_0$ and $h_0+g(n_c + \tilde {n}/2)$, where $h_0$ is the self-consistent mean field hamiltonian $h_0= V_{{\rm {ext}}}(r)-\mu +g{n_{c}}+2g\tilde{n}$, the TF condition $h_0 >> T$ ($T$ being the kinetic 
operator), if fulfilled for $n_c$ should also be satisfied for $\tilde{m}$. For this approximation to be consistent, 
$n_c$ and $\tilde{m}$ should vary on the same characteristic length, which is indeed the case as we will show later.

Before proceeding further, it is important to notice at this point that a kinetic-like term of the thermal cloud does 
not appear explicitly in the equations but is rather hidden in the third equation of (\ref{eq2}). Indeed, the kinetic term of the thermal cloud is related to the second derivative of the anomalous density. Differentiating (\ref{eq2t}) yields a relation of the form:

\begin{equation}
\Delta \tilde{n}\sim \left( \nabla |\tilde{m}|\right) ^{2}-\left( \nabla |
\tilde{n}|\right) ^{2}+|\tilde{m}|\Delta |\tilde{m}|,  
\label{eq2r}
\end{equation}
which shows in particular that neglecting $\Delta \tilde{m}$ does not necessarily mean neglecting $\Delta \tilde{n}$. That is precisely the recipe that we shall adopt below.

With this finite temperature prescription, the static equations now write

\begin{equation}
\begin{array}{rl}
& \left( V_{{\rm {ext}}}(r)-\mu +g{n_{c}}+2g\tilde{n}\right) \Phi +g\tilde{m}
\Phi ^{\ast }=0, \\ 
& \tilde {m}^{\ast }\Phi ^{2}-\tilde{m}{\Phi ^{\ast }}^{2}=0, \\ 
& \left( V_{{\rm {ext}}}(r)-\mu +2gn+{\frac{g}{4V}}(2V\tilde{n}+1)\right) 
\tilde{m}
+{\frac{g}{4V}}(2V\tilde{n}+1)\Phi ^{2}=0,
\end{array}
\label{eq2s}
\end{equation}
These equations are naturally gapless and satisfy the Hugenholtz-Pines theorem\cite{HF}. Indeed, owing to the second equation in (\ref{eq2s}), one can easily show that at zero momentum, the relation 
$\mu =g(n+\tilde{n}-|\tilde{m}|)$ is clearly satisfied without adding further assumptions, as is usually performed\cite{HF}.

In order to solve these equations, we may distinguish two different situations. The first one is for $T=0$. When all the atoms are condensed, $\tilde{n}=\tilde{m}=0$, and ${{n_{c}}}$ equals the total density $n$ of the gas. Omitting the trivial solution with ${{n_{c}}}=0$, one may take into account just the first equation in (\ref{eq2s}), since we consider a gas without a quantum cloud. Indeed, within the present set of equations, it is an approximation (although justifiable) to ignore the quantum depletion at $T=0$. The last two equations in (\ref{eq2s}) become therefore meaningless, and we are left with a simple expression for the condensate density 

\begin{equation}
{{n_{c}}}(r)=-\xi (r)={\frac{1}{g}}\left( \mu -V_{{\rm {ext}}}(r)\right) .
\label{eq4}
\end{equation}
Upon defining the oscillator length $a_{0}=(\hbar /m\omega _{0})^{1/2}$ and the s-wave scattering length 
$a=mg/4\pi \hbar ^{2}$, we obtain for a spherical trapping potential 
$V_{{\rm {ext}}}(r)={\frac{1}{2}}m\omega_{0}^{2}r^{2}$, the condensate radius $R$ and the reduced chemical potential 
$\nu _{0}=\mu /{{\frac{1}{2}}}\hbar \omega _{0}$ for a gas of $N$ bosons as

\begin{equation}
\frac{R}{a_0}=\left( 15N\frac{a}{a_0}\right) ^{1/5},  
\label{eq5}
\end{equation}

\begin{equation}
\nu _{0}=\left( 15N\frac{a}{a_0}\right) ^{2/5}.  
\label{eq55}
\end{equation}
The preceding expressions show that the spreading of the condensate depends essentially on the balance between the self-interactions and the trapping potential. These results have also been obtained by many other authors, see e.g. \cite{GR,ST,BM05}.

When $0\leq T<T_{{\rm {BEC}}}$, we have of course ${{n_{c}}}\neq 0$ and $\tilde{n}\neq 0$. Let us introduce the parametrization $2V\tilde{n}+1=\sqrt{I}\cosh {\sigma }$, $2V|\tilde{m}|=\sqrt{I}\sinh {\sigma }$, which automatically endows the relation (\ref{eq2t}). Then, from the third equation in (\ref{eq2s}), one obtains a simple equation for $X=e^{\sigma }$:

\begin{equation}
3X^{4}-4X^{2}+1+\frac{4Vn_{c}}{\sqrt{I}}\left( X^{2}-3\right) X=0,
\label{eq7}
\end{equation}
from which one extracts $\tilde{n}$ and $|\tilde{m}|$ as functions of $n_c$. Next, one uses these expressions in the first equation (\ref{eq2s}) to get the condensate density

\begin{equation}
{n_c}(r)=-\xi (r)-{\frac{1}{V}}\left( \frac{X+3X^{-1}}{4}\sqrt{I}-1\right) .  
\label{eq66}
\end{equation}
What is remarkable is that the sole acceptable solution of equation (\ref{eq7}) is a bounded function of $\eta=Vn_{c}/\sqrt{I}$. It is represented on figure 1.

\begin{figure}[h]

\begin{center}
\includegraphics [scale=1.]{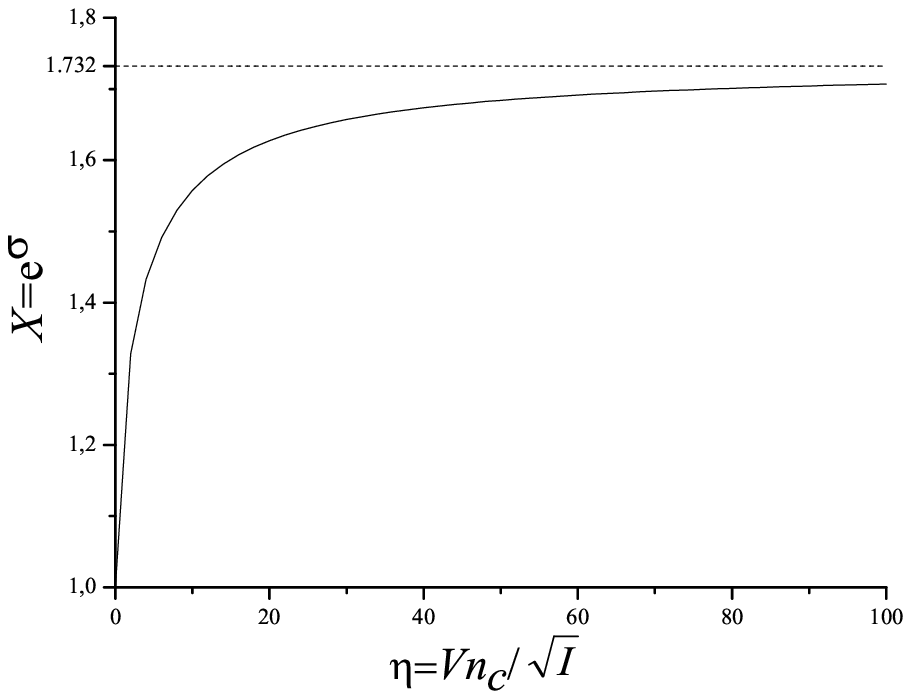}
\end{center}
\end{figure}

\vskip .7cm

\bc {\bf Fig.1:} Solution of Eq.(\ref{eq7}) vs. the dimensionless condensate density.\ec

%\hfill
%\begin{figure}[h]
%\begin{center}
%\includegraphics[scale=.6]{fig2.eps}
%\end{center}
%\end{figure}

Due to this behavior, one can easily show that the quantity $\frac{X+3X^{-1}}{4}$ which appears in (\ref{eq66}) is almost independent of $n_{c}$ and becomes rapidly close to unity. Hence, one may safely approximate (\ref{eq66}) by
\begin{equation}
{n_c}(r)\simeq -\xi (r)-{\frac{1}{V}}\left(\sqrt{I}-1\right) .
\label{eq67}
\end{equation}
In fact, one can check that the relative error between the two expressions (\ref{eq66}) and (\ref{eq67}) is less than 1\%. Finally, since $\sqrt{I}$ does not depend on space, the result (\ref{eq67}) shows that the finite temperature correction to the Thomas-Fermi profile (\ref{eq4}) is simply a space-independent (but temperature dependent) shift. The condensate density finally writes in the suggestive form
\begin{equation}
{n_c}(r)=\frac{V_{{\rm {ext}}}(R)-V_{{\rm {ext}}}(r)}{g},  
\label{eq8}
\end{equation}
which is formally the zero temperature TF profile. It is then easy to show that the condensate radius $R$ takes also a simple form
\begin{equation}
\frac{R}{a_0}=\left( 15N_{c}\frac{a}{a_0}\right) ^{1/5},  
\label{eq9}
\end{equation}
but now, it is the number of condensed atoms $N_{c}$ which is involved and not the total number of atoms. Hence, our finite temperature prescription for the TF approximation provides a natural extension of the zero temperature expressions, since the Thomas-Fermi parameter is now $N_{c}a/a_{0}$ instead of $Na/a_{0}$. The same conclusion may be drawn for the chemical potential which now writes 
\begin{equation}
\nu =\left(15N_c\frac{a}{a_0}\right)^{2/5} + 6\pi \frac{a}{a_0} (\sqrt{I}-1)\left(15N_c\frac{a}{a_0}\right) ^{-3/5}.
\label{eq667}
\end{equation}

In order to apprehend better these results, let us compute the remaining unknown quantities, such as the non condensed and the anomalous densities. To this end, and in order to obtain tractable expressions, we find it more convenient to use the simple fit
\begin{equation}
X=\frac{\sqrt{3}\eta +2/3}{\eta +2/3} ,  
\label{eq68}
\end{equation}
instead of the full analytical solution of equation (\ref{eq7}) which reproduces correctly the solution $X$ plotted in figure 1 with a residual error less than 0.1\%. Upon rewriting equation (\ref{eq8}) in the form $\eta=\eta _{0}(1-x^{2})$, with an obvious definition of $\eta _{0}$, we readily get
\begin{equation}
\tilde{n}(x)=\frac{1}{2V}\left\{ \frac{\sqrt{I}}{2}\left( \frac{\sqrt{3}\eta_{0}(1-x^{2})+{2/3}}{\eta _{0}(1-x^{2})+{2/3}}
+\frac{\eta _{0}(1-x^{2})+{2/3}}{\sqrt{3}\eta _{0}(1-x^{2})+{2/3}}\right) -1\right\} ,  
\label{eq10}
\end{equation}

\begin{equation}
\left\vert \tilde{m}\right\vert (x)=\frac{1}{2V}\frac{\sqrt{I}}{2}\left( 
\frac{\sqrt{3}\eta _{0}(1-x^{2})+{2/3}}{\eta _{0}(1-x^{2})+{2/3}}-\frac{\eta_{0}(1-x^{2})+{2/3}}{\sqrt{3}\eta _{0}(1-x^{2})+{2/3}}\right) {,}
\label{eq11}
\end{equation}
where $x$ is the reduced radial distance $x=r/R$. 

%\begin{figure}[h] 
%\begin{center}
%\includegraphics[scale=.7]{fig3.eps} 
%\end{center}
%\end{figure} 
%\hfill
%\begin{figure}[h] 
%\begin{center}
%\includegraphics[scale=.7]{fig4.eps} 
%\end{center} 
%\end{figure} 

In order to obtain more quantitative results, one must determine $N_{c}$ by using the normalization condition. We get easily the relation
\begin{equation}
1+2N=2N_{c}+\sqrt{I}g(s),  
\label{eq12}
\end{equation}
where
\begin{equation}
g(s)=\frac{2}{\sqrt{3}}+(\sqrt{3}-1)s\left\{ 1-\frac{3}{2}\sqrt{s+1}{\rm arc}\tanh \frac{1}{\sqrt{s+1}}
+\frac{1}{2}\sqrt{\frac{s}{\sqrt{3}}+1}{\rm arc}\tanh \frac{1}{\sqrt{\frac{s}{\sqrt{3}}+1}}\right\} ,  
\label{eq13}
\end{equation}
with $s=4\sqrt{I}/15N_{c}$. But since the function $g(s)$ satisfies $1\leq g(s)\leq 2/\sqrt{3}$, the equation (\ref{eq12}) is approximately solved to yield, to a very good accuracy, the simple result
\begin{equation}
N_{c}\simeq N-\frac{\sqrt{I}-1}{2}.  
\label{eq144}
\end{equation}
All the unknown quantities may now be determined in terms of $N$ and $\sqrt{I}$ alone. The corresponding results will be discussed in the next section.

\setcounter{equation}{0}

\section{Results and Discussions}
First of all, the condensate radius (\ref{eq9}) may be written as
\begin{equation}
R=R_{{\rm {TF}}}\left( \frac{N_{c}}{N}\right) ^{1/5},  
\label{eq21}
\end{equation}
where $R_{{\rm {TF}}}$ is the zero temperature result given by equation (\ref{eq5}). Figure 2 represents the condensate radius (in units of $R_{{\rm {TF}}}$) as a function of the condensate fraction and we notice in particular the compression of the condensate when reducing $N_{c}/N$ (that is increasing the temperature). This effect is by now a well established experimental result \cite{gerbier} and is attributed to the thermal cloud. The same effect of compression is observed on figure 3 for the central density $n_{c}(r=0)$ but it is more pronounced due to the power law of $2/5$ (see \ref{eq8}) instead of $1/5$ for the condensate radius. To be more precise, let us choose generic values for the number of atoms and the interaction strength ($N=10^{5}$ and $a/a_{0}=0.5\,10^{-3}$) and plot the various densities (in units of the oscillator volume $a_{0}^{3}$) versus the radial distance (in units of 
$\ R_{{\rm {TF}}}=3.758a_{0}$) for a condensate fraction ranging from $5\%$ up to $60\%$.

\begin{figure}[h] 
\begin{center}
\includegraphics[scale=1.]{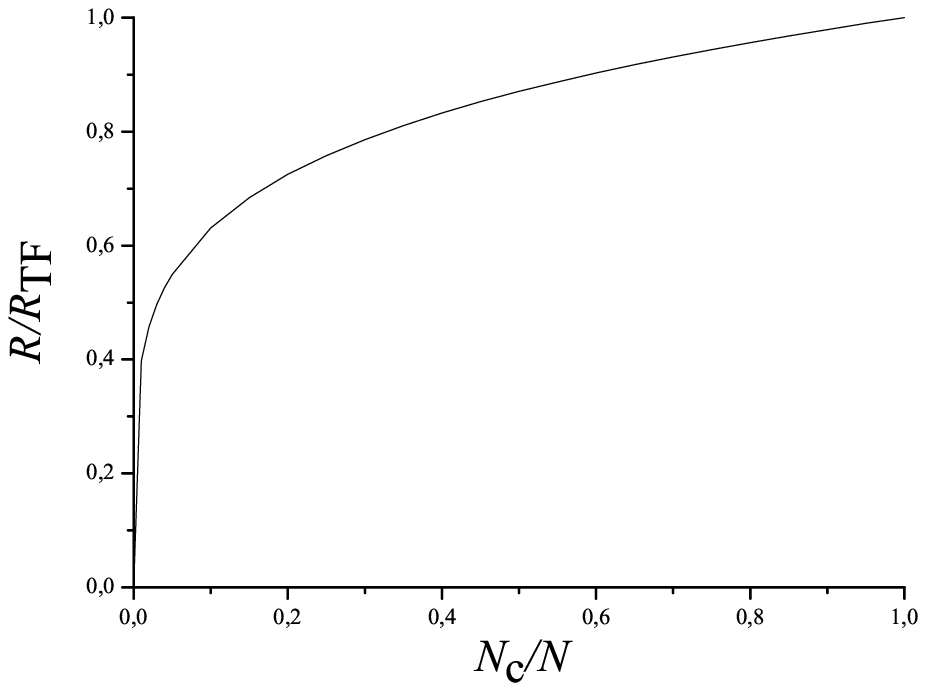} 
\end{center}
\end{figure}
%\vskip .7cm
\bc {\bf Fig.2:} Condensate radius vs. the condensate fraction.\ec
\vskip 2.cm
%\hfill
\begin{figure}[h] 
\begin{center}
\includegraphics[scale=1.]{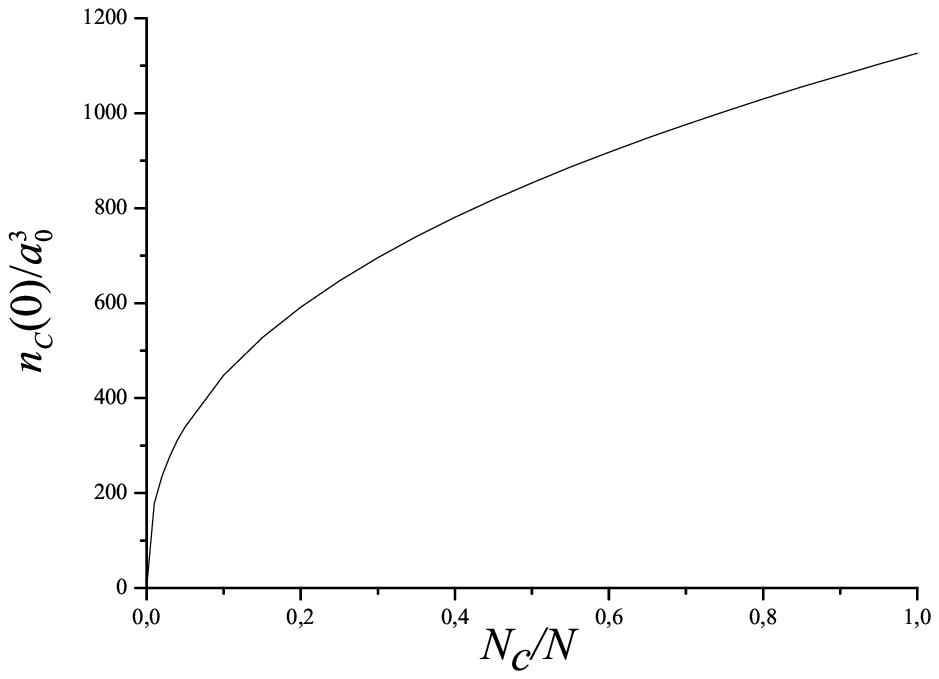} 
\end{center}
\end{figure}
\bc {\bf Fig.3:} Central density vs. the condensate fraction.\ec

The figure 4 shows typical Thomas-Fermi profiles for the condensate density, even for low condensate fraction. This is of course what one may expect on general grounds in the TF regime. Moreover, the effect of compression of the condensate is also clearly visible here.
\begin{figure}[h] 
\begin{center} 
\includegraphics[scale=1.]{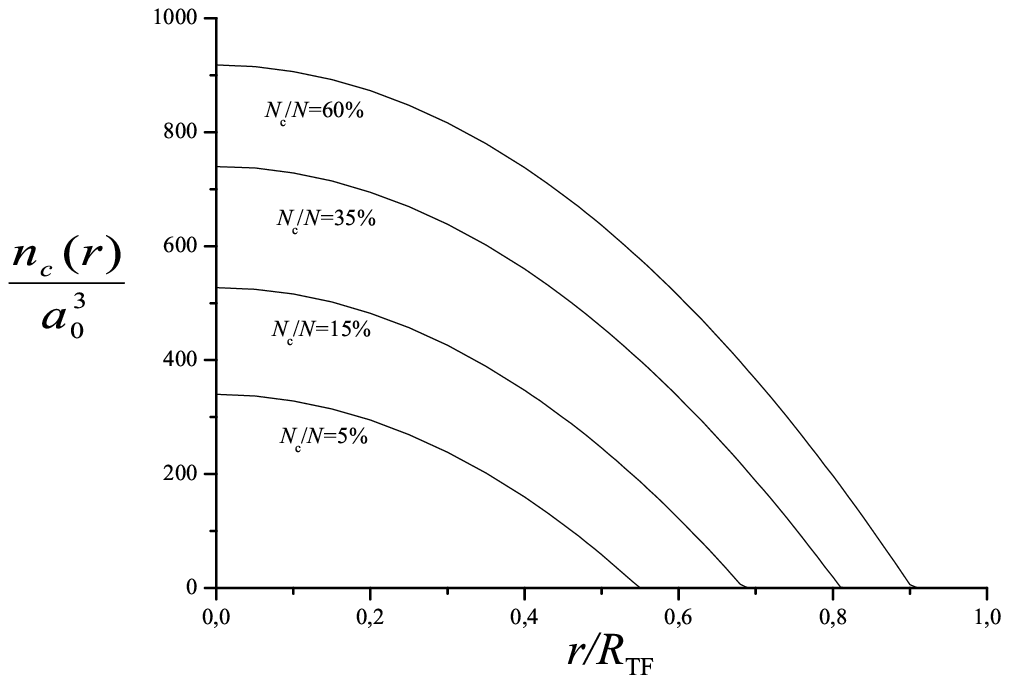} 
\end{center} 
\end{figure} 
\bc {\bf Fig.4:} Condensate density (in units of $a_0^{-3}$) vs. the radial distance (in units of $R_{\rm TF}$, see text) for $N=10^{5}$ and $a/a_0 =0.510^{-3}$.\ec
The non condensate density (from which we have subtracted a constant $\tilde{n}(R)$ for clarity) is plotted on figure 5 with the same units as before.

\begin{figure}[h] 
\begin{center} 
\includegraphics[scale=1.]{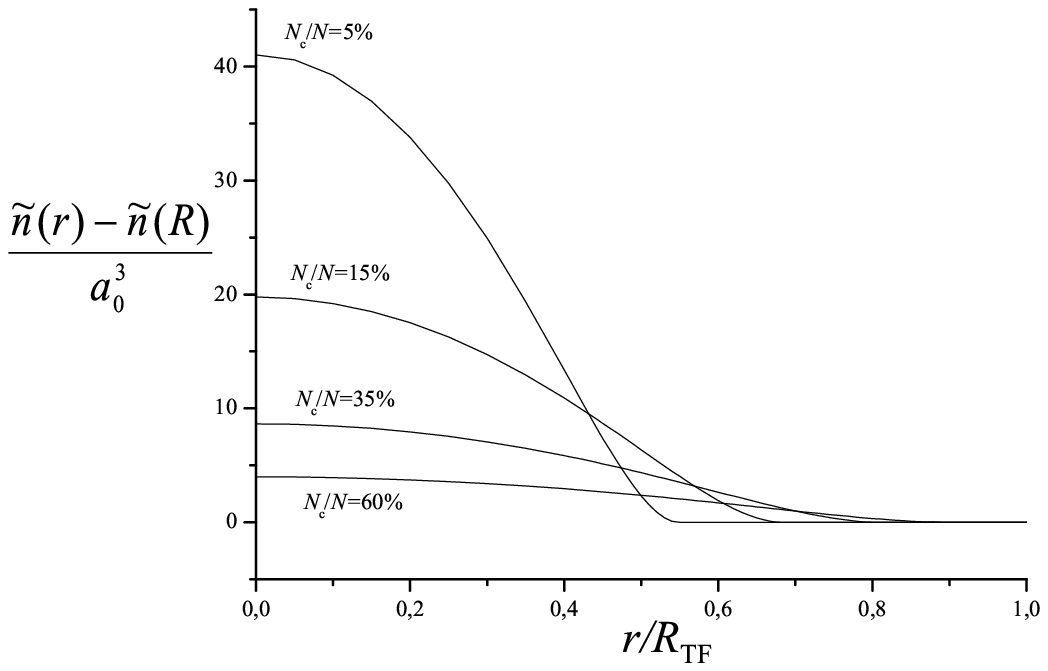} 
\end{center} 
\end{figure} 
\bc {\bf Fig.5:} Non condensate density (in units of $a_0^{-3}$) vs. the radial distance (in units of $R_{\rm TF}$) for $N=10^{5}$ and $a/a_0 =0.510^{-3}$.\ec

We notice that when increasing the condensate fraction, the thermal cloud tends to spread and flatten. This behavior is nicely comparable to the figure 4 of Ref.\cite{gerbier} (we have indeed deliberately chosen the same values for the condensate fraction. See however the remarks in sect. V of \cite{gerbier}\footnote{Although, in order to compare exactly with the experimental results, one must include an overall scale factor due to the finite ballistic expansion time.}). Furthermore, the non condensate density is peaked at the center of the trap for the whole range of the condensate fraction that we have considered. This is also confirmed by the experimental results of Gerbier {\it et al.} On the other hand, the full self-consistent HFB calculations\cite{HU00} predict that the non condensate density is sharply peaked at the edge of the condensate, having therefore a ''hole'' at the center of the trap. It seems however that this structure would shift toward the center with increasing atom number, but this behavior has not yet been checked\cite{HUP}. 

Nevertheless, we do understand in our case that neglecting the kinetic-like terms entails overestimating the (effective) repulsive self-interactions which tend to make the profiles more uniform. This overall effect is also observed on the anomalous density (fig. 6). Even if this quantity has not yet been measured experimentally, it is interesting to notice that our calculations predict a very simple and yet intuitive behavior as well which has been confirmed elsewhere\cite{HU00}.

On the other hand, the thermal cloud takes on a (small but) finite value for $r\geq R$. Even if this behavior is less intuitive, it is not very surprising since we do know that neglecting the second order derivatives amounts to making a cut of the densities at the boundaries. It is indeed a limitation of the TF approximation as a whole at the boundaries\cite{dalfovo}. The tail should be reproduced when one reinjects the second derivatives in the equations. In fact, it has been shown in the context of the full HFB numerical calculation\cite{HU00} that this is indeed the case.

\begin{figure}[h] 
\begin{center} 
\includegraphics[scale=1.]{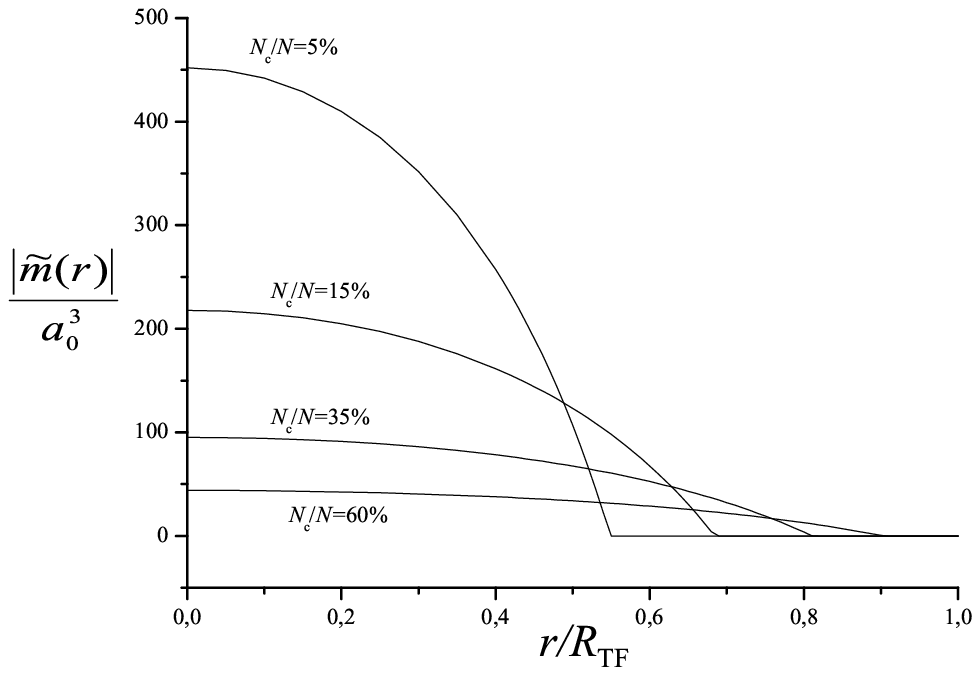} 
\end{center} 
\end{figure} 

\bc {\bf Fig.6:} Anomalous density (in units of $a_0^{-3}$) vs. the radial distance (in units of $R_{\rm TF}$) for $N=10^{5}$ and $a/a_0 =0.510^{-3}$.\ec
Finally, we observe on the figures 4, 5 and 6 that $n_c$, $|\tilde{m}|$ and $\tilde{n}$ vary on the same length scale ($R$) which justifies {\it a posteriori} our previous assumption.

\setcounter{equation}{0}{}
\section{Concluding Remarks}
We present in this paper a finite temperature analysis of the HFB equations in the Thomas-Fermi limit for a gas of bosons in a harmonic trap. At zero temperature, we obtain familiar expressions for the chemical potential and the condensate radius. The standard Thomas-Fermi profile for the condensate density is also recovered.

At finite temperatures and below the transition, since there are two phases, one should provide a prescription for the TF limit. We propose such a recipe (maybe the simplest) which consists in neglecting the second order derivatives of the condensate density and the anomalous density. The underlying idea is that, although the anomalous density is necessary for the coherence of the equations, it is believed to be a very small and a very smooth quantity. 
We therefore obtain analytical expressions for the condensate density, the condensate radius, the chemical potential and the central density as functions of the condensate fraction (or the temperature). Our expressions appear as natural extensions of the zero temperature TF limit.

Most importantly, we derive quite simple expressions for the non condensate density and for the anomalous density, which we plot as functions of the condensate fraction and draw many conclusions. First of all, the condensate profile is almost of the TF shape of which the spatial extension and the heights are controlled by $N_{c}/N$. Furthermore, the compression of the condensate by the thermal cloud with increasing temperature is clearly visible. On the other hand, the non condensate density profile is qualitatively consistent with the temperature dependence observed in recent experiments. In particular, the non condensate density is not peaked at the edge of the condensate (as predicted by the HFB calculations) but has a broad maximum at the center of the trap. The thermal cloud tends to spread and flatten with increasing temperature. The calculated anomalous density, although not yet observed experimentally, shows also a very intuitive behavior; it has a broad maximum at the center of the trap and vanishes at the boundaries. The tendency to spreading and flattening with increasing temperature is also observed here.
 
At the borders of the trap (where the condensate density vanishes) and for a given temperature, the non condensate density takes on a finite value which is a quite abrupt behavior. Although this meets the fact that the thermal cloud is actually surrounding the condensate, it is by no means conclusive. But this is a known shortcoming of the TF approximation as a whole since it breaks down at the boundaries of the condensate. Indeed, reinjecting the second derivatives of the densities will entail a more physical behavior. 

These results are quite satisfactory considering the drastic simplifications that occur, as compared for instance to the full self-consistent HFB calculations. Indeed, even if our overall density profiles are smoother and broader than what they should be, we know precisely the reasons and how to incorporate gradually what is missing.
 
It is important to emphasize that the self-consistent HFB equations obviously cannot be easily handled for high atom numbers. Hence, our equations may provide a simple tool to gain new insights in situations where these techniques become computationally consuming. One may make contact with the HFB calculations by taking into account the kinetic terms. That is precisely what we are actually performing in the semi-classical limit. The preliminary results show a qualitative and quantitative improvement of the predictions. More details are postponed to a forthcoming paper \cite{BS}. 

We are grateful to P. Schuck and Y. Castin for fruitful discussions and a careful reading of the manuscript. Special thanks to the members of the Groupe de Physique Th\'{e}orique, IPN-Orsay-France, where part of this work has been done.

\end{document}